\documentclass[aps,prl,superscriptaddress,preprintnumbers,twocolumn,groupedaddress,showpacs]{revtex4}
\usepackage{graphicx}
\usepackage{enumerate}

\def\beq{\begin{equation}}
\def\eeq{\end{equation}}
\def\bea{\begin{eqnarray}}
\def\eea{\end{eqnarray}}
\newcommand{\cO}{{\mathcal O}}
\newcommand{\nn}{\nonumber}
\newcommand{\eezzr}{$e^+e^- \to Z Z\gamma~$}
\newcommand{\eezh}{$e^+e^- \to Z H$}
\newcommand{\GeV}{{\rm GeV}}
\newcommand{\TeV}{{\rm TeV}}
\newcommand{\MeV}{{\rm MeV}}
\begin{document}

\title{  Precision study on $ZZ\gamma$ production including $Z$-boson leptonic decays at the ILC }
\author{ Zhang Yu$^a$, Duan Peng-Fei$^a$, Ma Wen-Gan$^b$, Zhang Ren-You$^b$ and Chen Chong$^b$ \\
{\small  $^a$ City College, Kunming University of Science and Technology, }  \\
{\small  Kunming, Yunnan 650051, People's Republic of China}  \\
{\small  $^b$ Department of Modern Physics, University of Science and Technology of China (USTC),}  \\
{\small   Hefei, Anhui 230026, People's Republic of China}}

\begin{abstract}
We report on the precision predictions for the \eezzr process including $Z$-boson leptonic decays at the ILC in the standard model (SM). The calculation includes the full next-to-leading (NLO) electroweak (EW) corrections and high order initial state radiation (h.o.ISR) contributions. We find that the NLO EW corrections heavily suppress the LO cross section, and the h.o.ISR effects are notable near the threshold while become small in high energy region. We present the LO and the NLO EW+h.o.ISR corrected distributions of the transverse momenta of final $Z$-boson and photon as well as the $Z$-pair invariant mass, and investigate the corresponding NLO EW and h.o.ISR relative corrections. We also study the leptonic decays of the final $Z$-boson pair by adopting the {\sc MadSpin} method where the spin correlation effect is involved. We conclude that both the h.o.ISR effects and the NLO EW corrections are important in exploring the $ZZ\gamma$ production at the ILC.
\end{abstract}

\pacs{12.15.Lk, 12.38.Bx, 14.70.Hp, 14.70.Bh}

\maketitle

\vskip 5mm
\section{Introduction}
\label{Sec:intro}
The standard model (SM) is the most successful particle physics model until now, since its theoretical predictions are consistent with high energy experimental results excellently. In the SM, electroweak symmety breaking (EWSB) is achieved by introducing the Higgs mechanism, which gives masses to the elementary particles and implies the existence of a SM Higgs boson. A giant step of particle physics was made with a new boson with mass of about $126~ {\rm GeV}$ observed by both ATLAS and CMS collaborations at the Large Hadron Collider (LHC) in July 2012 \cite{Aad:2012tfa,Chatrchyan:2012xdj}. The properties of this new boson are very well compatible with the SM Higgs boson but leave the room for new physics. One of the next important tasks is to investigate and measure the nature of this new particle, and finally to determine whether it is really the SM Higgs boson. The International Linear Collider (ILC) is an ideal machine to complete this task.

\par
The $H \rightarrow Z \gamma$ channel is a rare decay mode but remarkable to discover the nature of the Higgs boson. This channel does not occur at tree level but is induced by loop diagrams mediated by a heavy quark loop \cite{Cahn:1978nz} or a $W$-boson loop \cite{Bergstrom:1985hp}, just like the $H \rightarrow \gamma\gamma$ decay. The $H \rightarrow Z \gamma$ decay mode may provide hint for new physics beyond the SM, because if new particles circulate in the loop or $H$ is a non-SM scalar boson, the $H \rightarrow Z \gamma$ decay rate will be different from the SM prediction. At the ILC, the Higgs-strahlung, \eezh, is the predominantly process for Higgs production. Therefore, the most serous and irreducible background arises from the \eezzr process in the search of Higgs boson via $H \rightarrow Z \gamma$ decay channel at the ILC.

\par
The multiple gauge boson production plays an important role in probing the gauge self-couplings and would be helpful for verification of the non-Abelian gauge structure of the SM. The direct measurements of the quartic gauge boson couplings (QGCs), which can provide a connection of the mechanism of EWSB, require the theoretical study of triple gauge boson production. In the last few years, the calculations up to the QCD next-to-leading order (NLO) on triple gauge boson production at hadron colliders in the SM have been completed \cite{Lazopoulos:2007ix,Hankele:2007sb,Campanario:2008yg,Binoth:2008kt,Bozzi:2009ig,Bozzi:2010sj,Baur:2010zf,Bozzi:2011wwa,Bozzi:2011en}. The NLO EW corrections to $WWZ$ and $WZZ$ productions at the LHC have been provided in Refs.\cite{Nhung:2013jta,Yong-Bai:2015xna}. The NLO EW corrections to $WWZ$, $ZZZ$, $WW\gamma$ and $Z\gamma\gamma$ productions at the ILC including the h.o.ISR contributions are calculated in Refs.\cite{JiJuan:2008nn,Wei:2009hq,Boudjema:2009pw,Chong:2014rea,Yu:2013dxa}.

\par
The $ZZ\gamma$ production at the ILC is not only an important background process to study the nature of Higgs boson, but also a signal process to explore the $ZZZ\gamma$ and $ZZ\gamma\gamma$ anomalous QGCs. These anomalous QGCs vanish at tree level in the SM and might provide a clean signal of new physics if any deviations from the SM predictions are observed. The effects of anomalous QGCs in $ZZ\gamma$ production at the LEP, ILC and CLIC were phenomenologically studied in Refs.\cite{Belanger:1992qh,Belanger:1999aw,Stirling:1999ek,Koksal:2014nua}. The theoretical leading order (LO) predictions in the SM were provided in Ref.\cite{Stirling:1999ek}, while the EW corrections to $e^+e^- \rightarrow ZZ\gamma$, which would be necessary to match the ILC experimental accuracy, are still missing.

\par
In this paper, we investigate the full NLO EW corrections and the h.o.ISR contributions to the $ZZ\gamma$ production at the ILC in the SM. The rest of the paper is organized as follows: In the following section we present the NLO EW and h.o.ISR analytical calculations for the \eezzr process. The numerical results and discussions are given in Section III. Finally, we give a short summary.

\vskip 5mm
\section{Calculation setup}
\label{Sec:setup}
In our calculation, we use the 't Hooft-Feynman gauge and apply {\sc FeynArts-3.7} package \cite{Hahn:2000kx} to generate automatically the Feynman diagrams. The corresponding amplitudes are subsequently reduced by using {\sc FormCalc-7.4} program \cite{Hahn:1998yk}. Because of the smallness of the electron mass, we neglect the contributions from the Feynman diagrams which involve the Higgs/Goldstone-electron-positron Yukawa couplings. We adopt a mixed scheme for the electromagnetic coupling, just as our previous work \cite{Chong:2014rea}, i.e., the couplings related to the external photons are fixed with $\alpha=\alpha(0)$ in the $\alpha(0)$-scheme and the others with $\alpha=\alpha_{G_\mu}=\frac{\sqrt{2}G_\mu M_W^2}{\pi}\left(1-\frac{M_W^2}{M_Z^2}\right)$ in the $G_\mu$-scheme. Then the LO and NLO EW corrected cross sections for $e^+e^- \rightarrow ZZ\gamma$ are of ${\cal O}(\alpha^2_{G_\mu}\alpha(0))$ and ${\cal O}(\alpha^2_{G_\mu}\alpha(0)^2)$, respectively.

\par
\subsection{Virtual correction}
The NLO EW correction to the $e^+e^- \rightarrow ZZ\gamma$ process consists of virtual loop correction and real photon emission correction. The virtual EW correction involves 1382 diagrams which can be classified into self-energy (64), triangle (722), box (507), pentagon (56) and counterterm (33) graph groups. The amplitude for these one-loop Feynman diagrams contains both ultraviolet (UV) and infrared (IR) singularities. We adopt the dimensional regularization scheme to regularize the UV divergences in loop integrals. The relevant renormalization constants are detailed in Refs.\cite{Ross:1973fp,Denner:1991kt} by using the on-mass-shell conditions. After performing the renormalization procedure, the UV divergences are canceled exactly. Since the logarithmic divergences contributed by light quarks to the photon wave-function renormalization constant have been absorbed in $\alpha_{G_\mu}$, we should subtract these contributions from the virtual correction in order to avoid double counting, i.e., the electric charge renormalization constant is given in the $G_\mu$-scheme as
\begin{eqnarray}
\delta Z_e^{G_{\mu}}= \delta Z_e^{\alpha(0)}-\frac{1}{2}\Delta r,
\end{eqnarray}
where $\delta Z_e^{\alpha(0)}$ is the electric charge renormalization constant in the $\alpha(0)$-scheme and $\Delta r$ comes from the weak corrections to muon decay \cite{Dittmaier:2001ay}. The IR singularities are regularized by using infinitesimal fictitious photon mass. After adding the contribution of real photon emission process, the IR singularities are canceled. Finally, a UV- and IR-finite EW correction can be obtained.

\par
We adopt the {\sc LoopTools-2.8} package \cite{Hahn:1998yk} for the numerical calculations of the scalar and tensor integrals, in which the $n$-point ($n\le 4$) tensor integrals are reduced to scalar integrals recursively by using Passarino-Veltman algorithm and the 5-point integrals are decomposed into 4-point integrals by using the method in Ref.\cite{Denner:2002ii}. In our previous work \cite{Chong:2014rea,Yu:2014cka}, we addressed the numerical instability originating from the small Gram determinant ($detG$) scalar $4$-point integrals \cite{Boudjema:2009pw}. In order to solve this problem, we developed the {\sc LoopTools-2.8} package and checked the results with those by using {\sc OneLoop} package \cite{vanHameren:2009dr} to verify the correctness of our codes.

\par
\begin{figure}[htbp]
\includegraphics[ scale = 1.0]{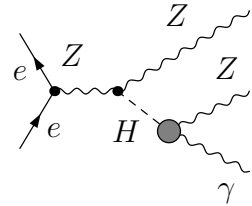}
\caption{ One-loop Feynman diagrams with internal on-shell Higgs boson for the $e^+e^- \rightarrow ZZ\gamma$ process.} \label{fig-res}
\end{figure}

The one-loop Feynman diagrams for $e^+e^- \rightarrow ZZ\gamma$ with possible Higgs resonance are demonstrated in Fig.\ref{fig-res}. The interference between the amplitudes for these one-loop Feynman diagrams and the LO diagrams leads to a propagator factor of $\frac{1}{(M_{Z\gamma}^2-M_{H}^2)}$, which is divergent in the vicinity of $M_{Z\gamma}^2 = M_{H}^2$. In order to regulate it, we replace $\frac{1}{(M_{Z\gamma}^2-M_{H}^2)}$ with $\frac{1}{(M_{Z\gamma}^2-M_{H}^2 +iM_H\Gamma_H)}$. The contribution of the imaginary part is so tiny that can be ignored in the total NLO EW correction.

\par
\subsection{Real photon emission correction}
Technically, to extract the IR singularities from the real photon emission correction and combine it with the virtual contribution, we employ the dipole subtraction (DS) method \cite{dipole-qed}. In the DS approach, the IR-finite real correction is obtained by subtracting an auxiliary function from the squared amplitude for the real photon emission process before integrating over phase space. Because the subtraction function have the same singularity structure as the squared amplitude, this phase space integration can be performed numerically. In order to obtain an unchanged result, the subtracted term is added again after analytical integration over the bremsstrahlung photon phase space. The dipole subtraction formalism is a process independent approach, which was first presented by Catani and Seymour for QCD with massless unpolarized partons \cite{Catani:1996jh,Catani:1996vz,Catani:2002hc} and subsequently was generalized to photon radiation off charged fermions with arbitrary mass by Dittmaier \cite{Dittmaier:1999mb}. In our  calculations, we follow the approach of Ref.\cite{Dittmaier:1999mb} and check the independence on the parameter $\alpha\in(0,1]$, which is introduced to control the size of dipole phase space \cite{Nagy:1998bb,Nagy:2003tz}.

\par
\subsection{High order initial state radiation}
The emission of photon collinear to the incoming electron or positron, i.e., initial state radiation (ISR), induces the collinear IR quasi-singularities due to the smallness of the electron mass. The virtual correction can partially cancel the ISR collinear IR quasi-singularities. The left ones would lead to large radiative corrections in the form $\alpha^n\log^n(m_e^2/Q^2)$ at the leading logarithmic (LL) level. In order to achieve an accuracy at the few $0.1\%$ level, the high order contributions from this part beyond NLO have to be taken into account. According to the mass factorization theorem, the LL initial state QED correction can be expressed as a convolution of the LO cross section with structure functions by using the structure function method \cite{Denner:2000bj,Beenakker:1996kt},
\bea
\label{eq:isr}
\int d\sigma_{\rm ISR-LL}&=&\int_0^1dx_1\int_0^1dx_2\Gamma_{ee}^{\rm LL}(x_1,Q^2)\Gamma_{ee}^{\rm LL}(x_2,Q^2)\nn\\
&&\times \int d\sigma(x_1p_{e^-},x_2p_{e^+}),
\eea
where $x_1$ and $x_2$ denote the momentum fractions carried by the incoming electron and positron just before the hard scattering, $Q^2$ is the typical scale at which the hard scattering occurs chosen as the colliding energy $\sqrt{s}$ in our calculation, and the LL structure functions $\Gamma_{ee}^{\rm LL}(x,Q^2)$ are detailed in Ref.\cite{Beenakker:1996kt} up to $\cO(\alpha^3)$. In summing the contribution from Eq.(\ref{eq:isr}) with the NLO EW corrected result, we must subtract the LO and one-loop contributions to avoid double counting. The explicit expression for the subtracted terms are presented in  Ref.\cite{Denner:2000bj}. In the following, the subtracted ISR effect is called the high order ISR (h.o.ISR) contribution beyond $\cO(\alpha)$. We define the summation of the NLO EW corrected cross section and the h.o.ISR contribution as the total EW corrected result.

\vskip 5mm
\section{Numerical results}
\par
\subsection{Input parameters and event selection criterion}
The relevant input parameters used in our calculation are taken as \cite{Agashe:2014kda}:
\begin{equation}\arraycolsep 2pt
\begin{array}[b]{lcllcllcl}
G_{\mu} &=& 1.1663787\times 10^{-5}~\GeV^{-2}, \\
\alpha(0)&=&1/137.035999074,\\
 M_{W} & = & 80.385~\GeV,  \qquad M_{Z}  =  91.1876~\GeV, \\
 m_e &=& 0.510998928~\MeV, \quad m_{\mu}=105.6583715~\MeV, \\
 m_{\tau}&=&1.77682~\GeV,\\
 m_u &=& 66~\MeV, \quad \qquad m_d = 66~\MeV, \\
 m_c&=&1.2~\GeV,\qquad m_s=150~\MeV, \\
 m_t&=&173.21~\GeV, \qquad m_b=4.3~\GeV,
\end{array}
\label{SMpar}
\end{equation}
where the current mass values of light quarks (all quarks except $t$-quark) can reproduce the hadronic contribution to the photonic vacuum polarization \cite{Jegerlehner:2001ca}. The Higgs boson mass is taken as $M_H=125.09~\GeV$ \cite{Aad:2015zhl}, and its decay width is obtained by using the HDECAY program \cite{Djouadi:1997yw}. The Cabibbo-Kobayashi-Maskawa (CKM) matrix is set to be unit matrix.

\par
There is only one photon in the final state of the process \eezzr at the LO, while at most two photons up to EW NLO. We apply the Cambridge/Aachen (C/A) jet algorithm \cite{Salam:2009jx} to the photon candidates. For a two-photon event originating from the real emission correction, we merge them into one new photon with four-momentum $p_{ij,\mu}=p_{i,\mu}+p_{j,\mu}$ when the separation $R = \sqrt{\Delta y^2 + \Delta \phi^2}$ of the two final photons satisfies the condition of $R<0.4$, where $\Delta y$ and $\Delta \phi$ are the differences of rapidity and azimuthal angle between the two photons, then we call this event as a "one-photon" event, otherwise it is considered as a "two-photon" event. In our calculations, we collect all the "one-photon" and "two-photon" events, and at least one photon is required to satisfy the constraints
\begin{eqnarray}\label{cut}
p^{\gamma}_T \ge 15~\GeV,~~~|y_{\gamma}|\le2.5~.
\end{eqnarray}
Thereby we exclude the inevitable infrared (IR) singularity in the LO calculation. In the "two-photon" event, when both the two photons pass the cuts in Eq.(\ref{cut}) we call the photon with the largest transverse energy as the leading photon and the another one as the sub-leading photon.

\par
\subsection{Total cross section}
\begin{figure}[htbp]
\includegraphics[angle=0,width=3.2in,height=2.4in]{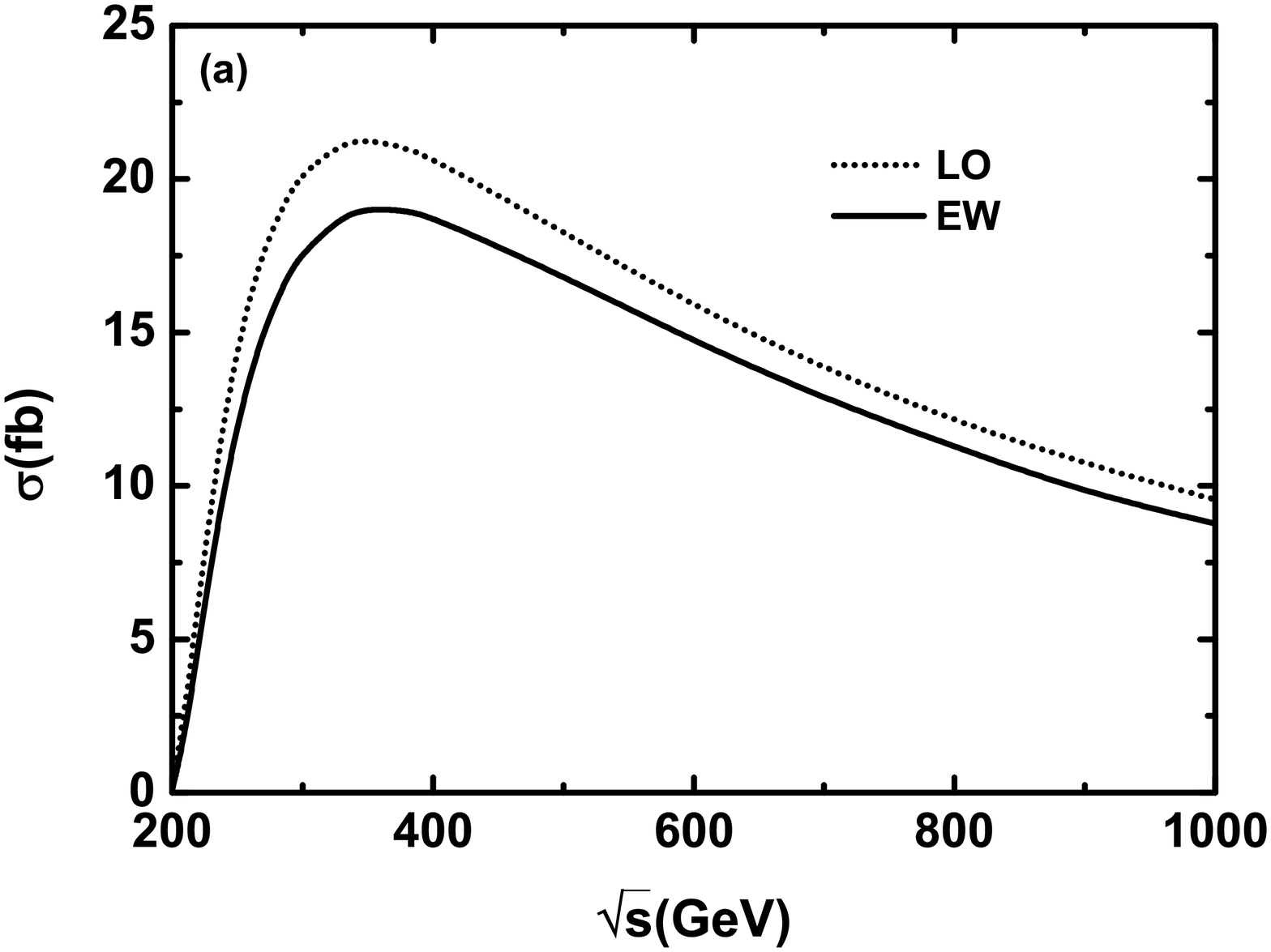}%
\hspace{0in}%
\includegraphics[angle=0,width=3.2in,height=2.4in]{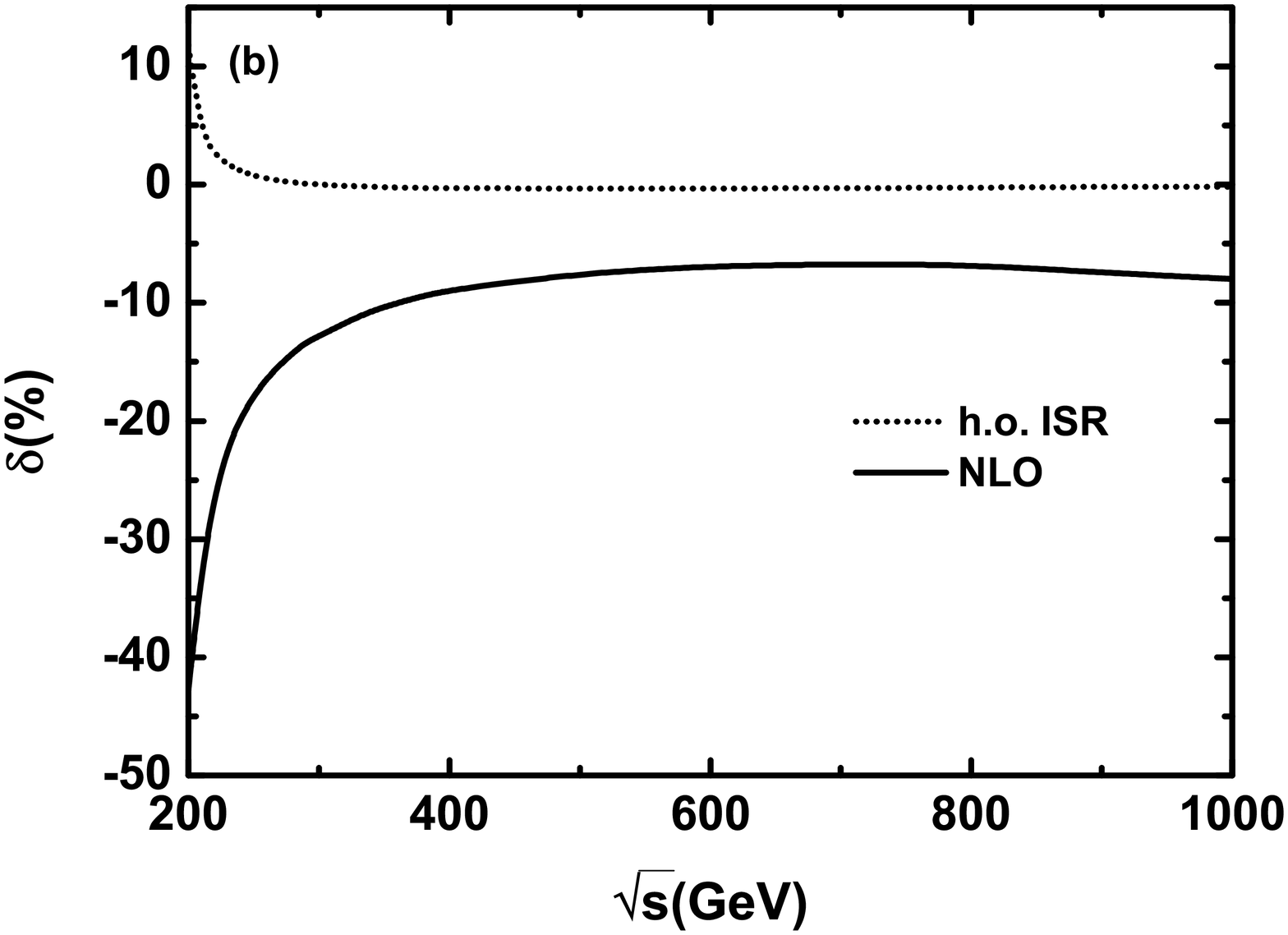}%
\hspace{0in}%
\caption{(a) The LO and total EW corrected cross sections ($\sigma_{{\rm LO}}$ and $\sigma_{{\rm EW}}$) for the \eezzr process as the functions of the colliding energy $\sqrt{s}$ at the ILC. (b) The corresponding NLO EW and h.o.ISR relative corrections ($\delta_{{\rm NLO}}$ and $\delta_{{\rm h.o.ISR}}$).} \label{fig-sqrts}
\end{figure}

In Fig.\ref{fig-sqrts}(a), we present the LO and total EW corrected integrated cross sections as the functions of the colliding energy $\sqrt s$ for the \eezzr process in the SM, and in Fig.\ref{fig-sqrts}(b) we show the corresponding h.o.ISR and NLO EW relative corrections, defined as $\delta\equiv\frac{\sigma-\sigma_{{\rm LO}}}{\sigma_{{\rm LO}}}$. From these figures, we find that all the LO and total EW corrected integrated cross sections are sensitive to the colliding energy, and they reach their maxima at the position of $\sqrt s\sim 350~\GeV$. The LO cross sections are suppressed by both the NLO EW and total EW correction in the whole plotted $\sqrt s$ region. From Fig.\ref{fig-sqrts}(b), we can see that the NLO EW relative corrections  near the threshold are very large. That is because for energies near the threshold the virtual EW corrections are enhanced by Coulombic photon exchange between the electron and positron in the initial state. When the photon momentum approaches to zero a so-called Coulomb singularity occurs. At high energy range the NLO EW relative correction is also notable and increases slowly with the increment of $\sqrt s$. That is due to the Sudakov logarithms from the virtual exchange of soft or collinear massive weak gauge bosons which dominate the weak corrections at high energies. As indicated in Fig.\ref{fig-sqrts}(b), the h.o.ISR effect beyond $\cO(\alpha)$ is significant near the threshold whose relative correction can reach $11.81\%$ at $\sqrt s = 200~\GeV$, but goes down and approaches the value lower than $1\%$ when $\sqrt s >250~\GeV$. In order to show the results more explicitly, we present some representative numerical results of the LO and total EW corrected cross sections, and the corresponding NLO EW and h.o.ISR relative corrections in Tab.\ref{tab1}.
\begin{table}
\center
\begin{tabular}{ccccc}
\hline
$\sqrt{s}(\GeV)$ & $\sigma_{{\rm LO}}({\rm fb})$ & $\sigma_{{\rm EW}}({\rm fb})$
& $\delta_{{\rm NLO}}(\%)$& $\delta_{{\rm h.o.ISR}}(\%)$ \\
\hline \hline
200   &0.12000(4)  & 0.08289(12)& -42.74   & 11.81\\
220   &6.212(2)    & 4.733(4)   & -26.43   & 2.62 \\
250   &14.428(4)   & 11.951(9)  & -17.95   & 0.78 \\
300   &20.180(4)   & 17.592(9)  & -12.84   & 0.01 \\
350   &21.288(4)   &19.040(9)   & -10.34   & -0.22\\
400   & 20.662(9)  & 18.756(15) & -8.93    & -0.30\\
500   & 18.255(9)  & 16.831(14) & -7.47    & -0.33\\
800   & 12.116(8)  & 11.265(13) & -6.78    & -0.25\\
1000  & 9.551(7)   & 8.771(13)  & -7.99    & -0.19\\
\hline  \hline
\end{tabular}
\caption{ \label{tab1}  The total LO cross section ($\sigma_{{\rm LO}}$), total EW corrected integrated cross sections ($\sigma_{{\rm EW}}$) and the corresponding NLO EW and h.o.ISR relative corrections ($\delta_{{\rm NLO}}$) and $\delta_{{\rm h.o.ISR}}$) for the \eezzr process.}
\end{table}

\par
\subsection{Kinematic distributions}
\begin{figure}[htbp]
\includegraphics[angle=0,width=3.2in,height=2.4in]{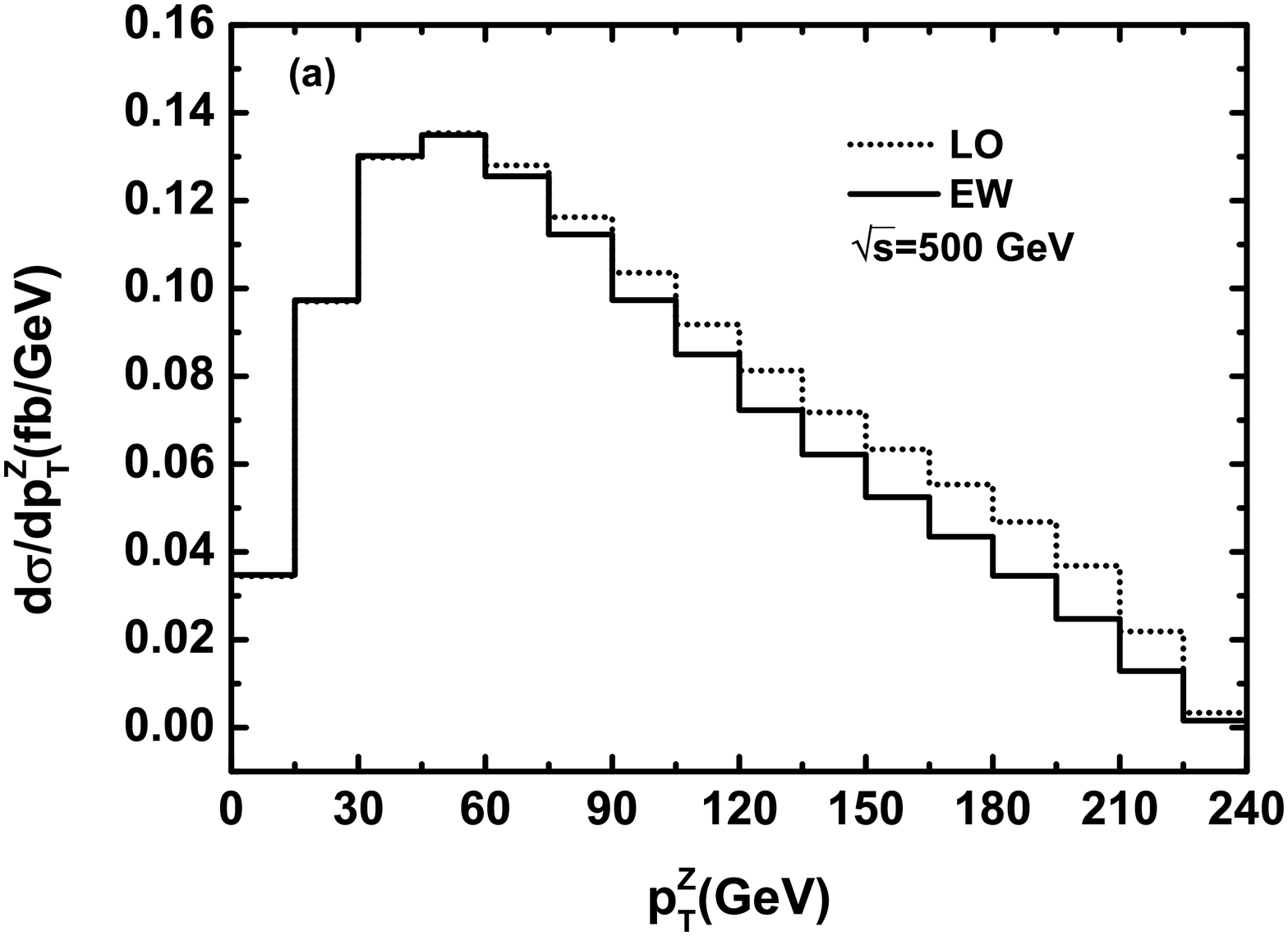}%
\hspace{0in}%
\includegraphics[angle=0,width=3.2in,height=2.4in]{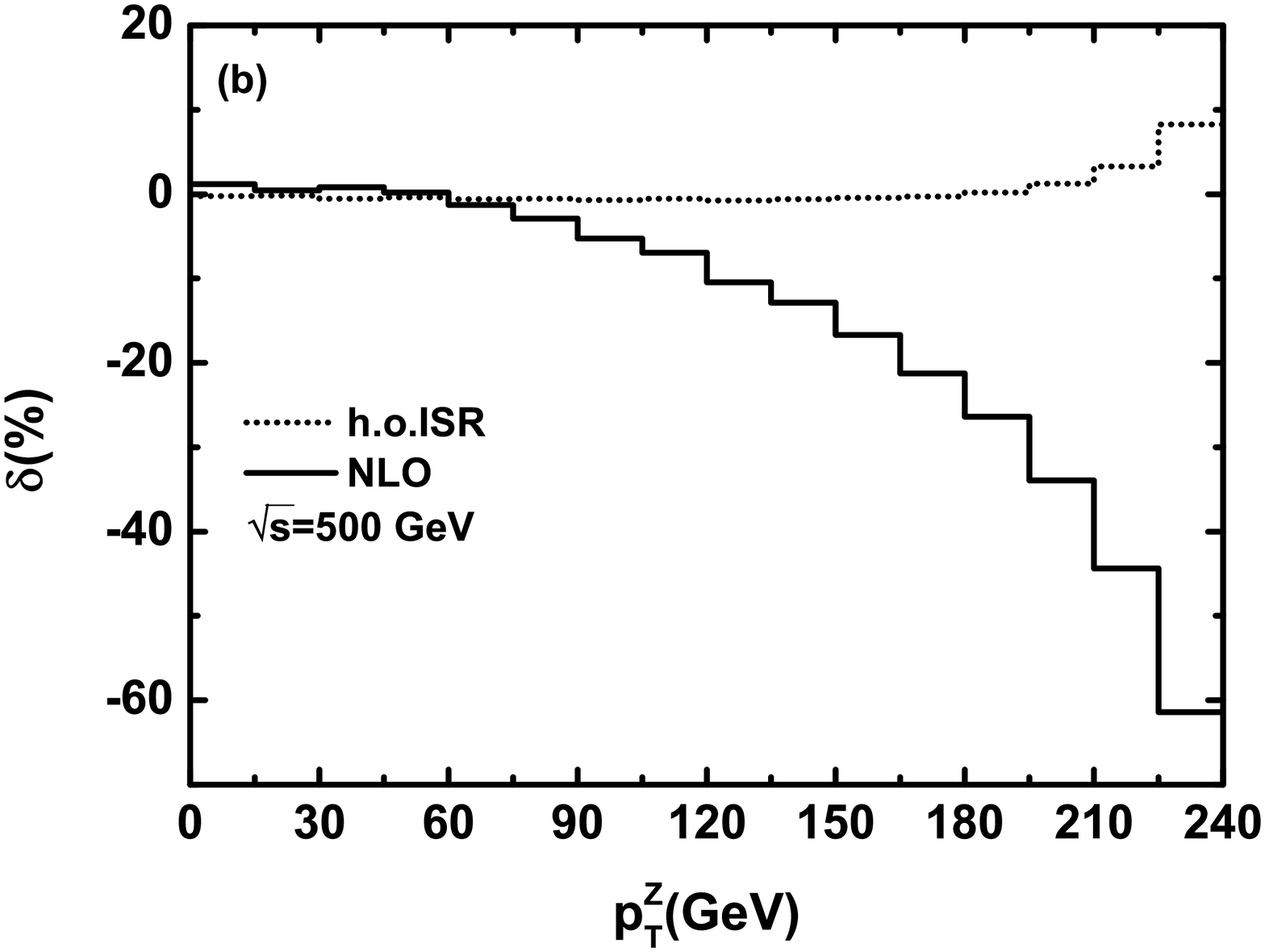}%
\hspace{0in}%
\caption{(a) The LO and total EW corrected transverse momentum distributions of final $Z$-boson at the $\sqrt s=500~\GeV$ ILC. (b) The corresponding NLO EW and h.o.ISR relative corrections. }\label{fig-ptz}
\end{figure}

\begin{figure}[htbp]
\includegraphics[angle=0,width=3.2in,height=2.4in]{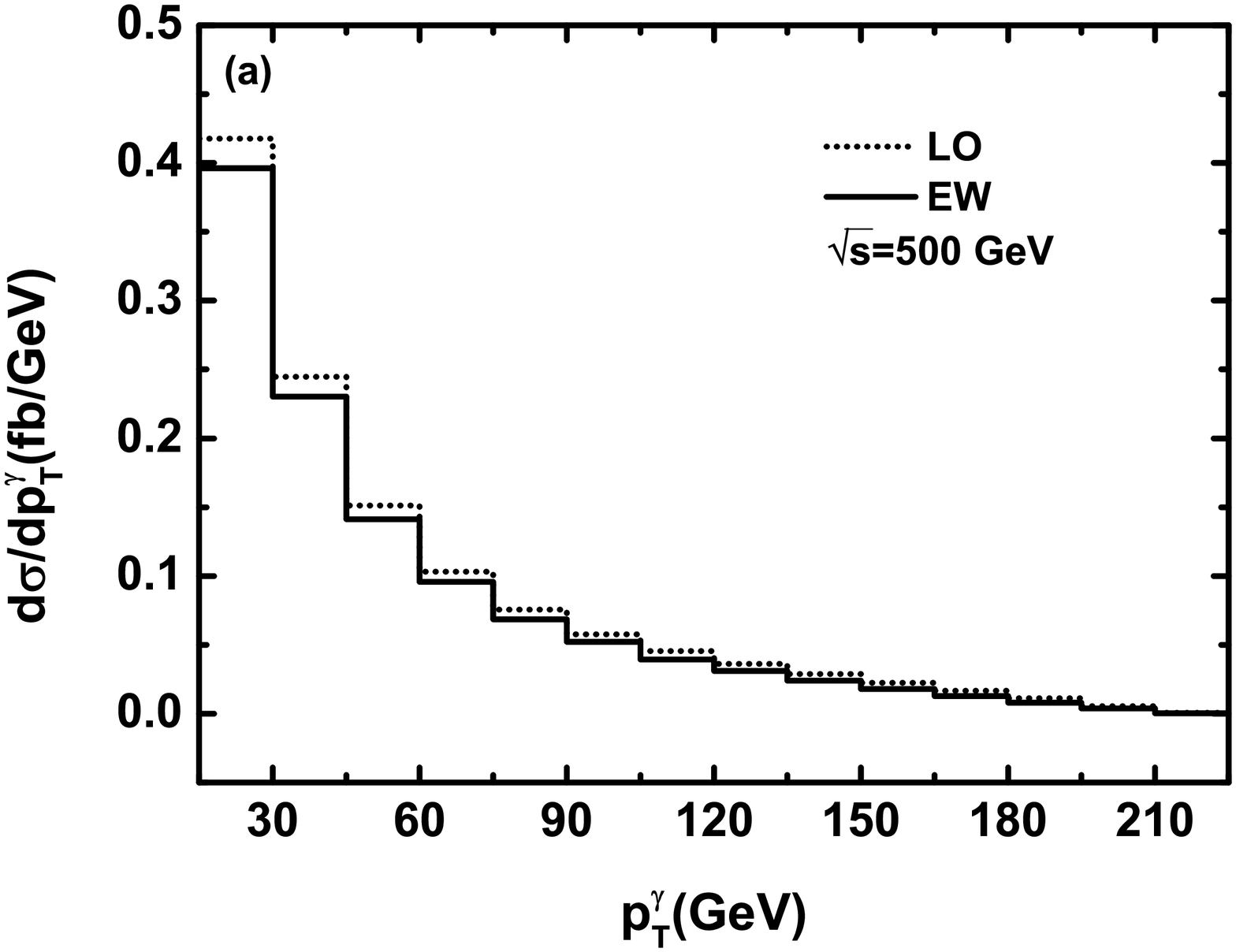}%
\hspace{0in}%
\includegraphics[angle=0,width=3.2in,height=2.4in]{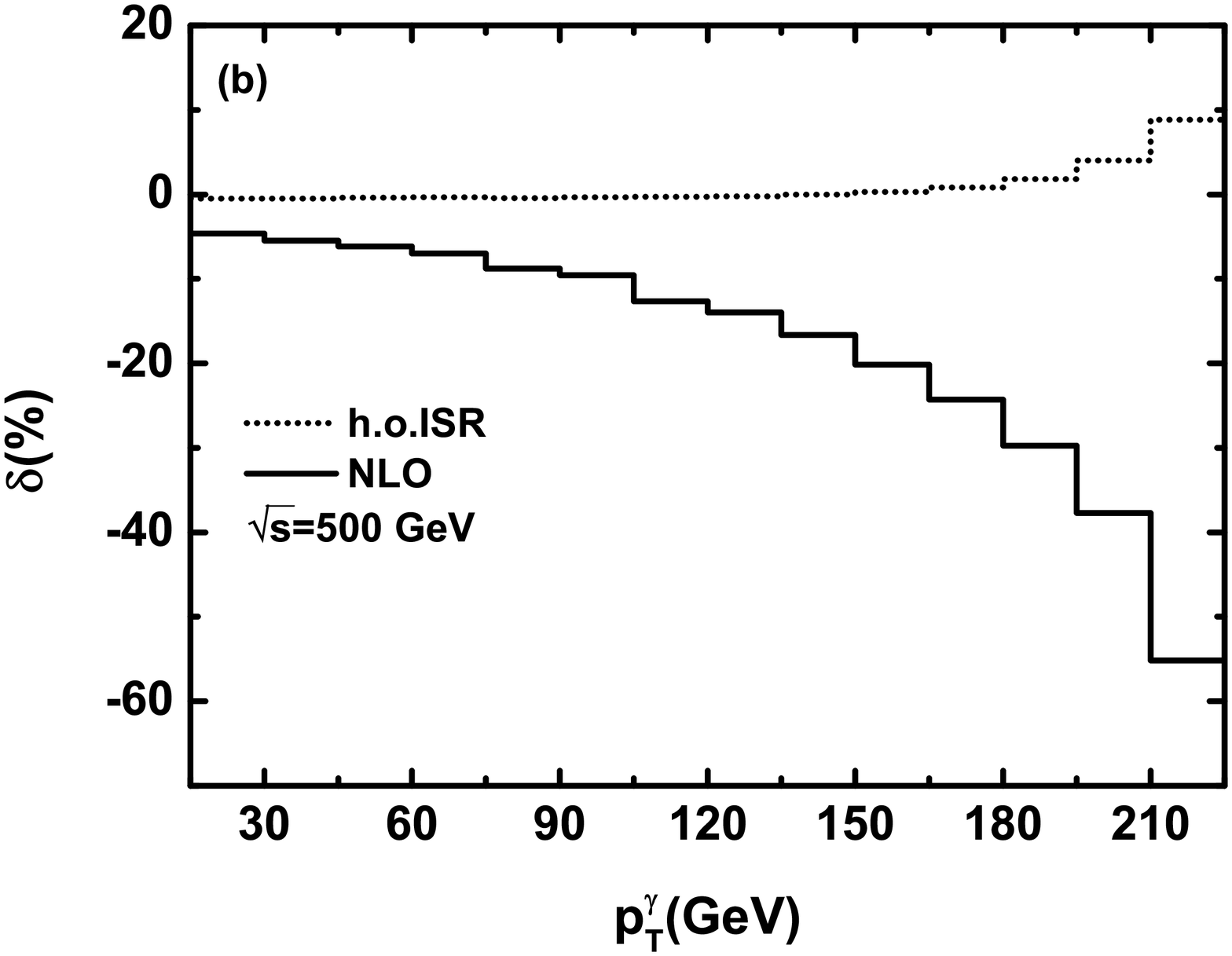}%
\hspace{0in}%
\caption{(a) The LO and total EW corrected transverse momentum distributions of final leading photon at the $\sqrt s=500~\GeV$ ILC. (b) The corresponding NLO EW and h.o.ISR relative corrections. } \label{fig-ptr}
\end{figure}

We discuss the kinematic distributions of final produced particles for the $ZZ\gamma$ production process in this subsection. The LO and total EW corrected transverse momentum distributions of the final $Z$-boson (i.e.,$\frac{d\sigma_{LO}}{dp_T^Z}$ and $\frac{d\sigma_{{\rm EW}}}{dp_T^Z}$) at the $\sqrt s=500~\GeV$ ILC are provided in Fig.\ref{fig-ptz}(a). There we pick the $p_T^Z$ of each of the two $Z$-bosons as an entry in this histograms, then the final result of the differential cross section is obtained by multiplying
factor $1/2$. From this figure, we can see that the LO differential cross sections are enhanced by the total EW correction when $p_T^Z\le45~\GeV$, while suppressed in the rest of plotted region.
The transverse momentum distributions of the leading photon are plotted in Fig.\ref{fig-ptr}(a).
It shows that both the LO and total EW corrected $p_T^\gamma$ distributions for the leading photon
decrease continuously with the increment of $p_T^\gamma$, and the LO $p_T^\gamma$ distributions are always
suppressed by the total EW correction. We also present the corresponding NLO EW and h.o.ISR relative correction distributions of $p_T^Z$ and $p_T^\gamma$ in Fig.\ref{fig-ptz}(b) and Fig.\ref{fig-ptr}(b), respectively. From these figures we can see that the h.o.ISR relative correction is usually very small, but when $p_T^Z$ ($p_T^\gamma$) approaches to 240 (225) GeV, the h.o.ISR relative correction can reach $8.3\%$ ($8.8\%$). Due to the Sudakov effect, the absolute size of the NLO EW relative corrections continuously grow up with the increments of $p_T^Z$ and $p_T^\gamma$ at high $p_T$ regions.

\par
\begin{figure}[htbp]
\includegraphics[angle=0,width=3.2in,height=2.4in]{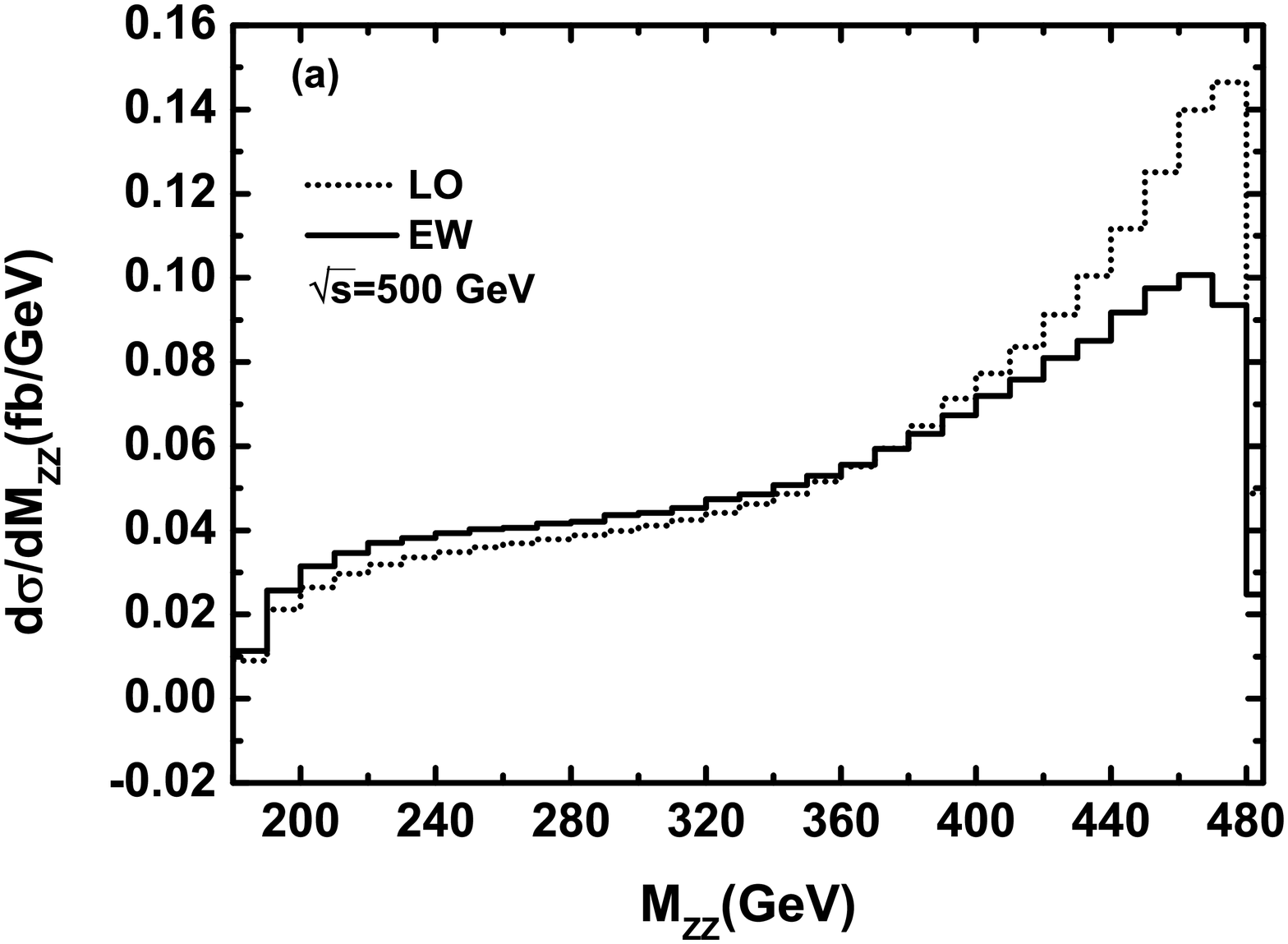}%
\hspace{0in}%
\includegraphics[angle=0,width=3.2in,height=2.4in]{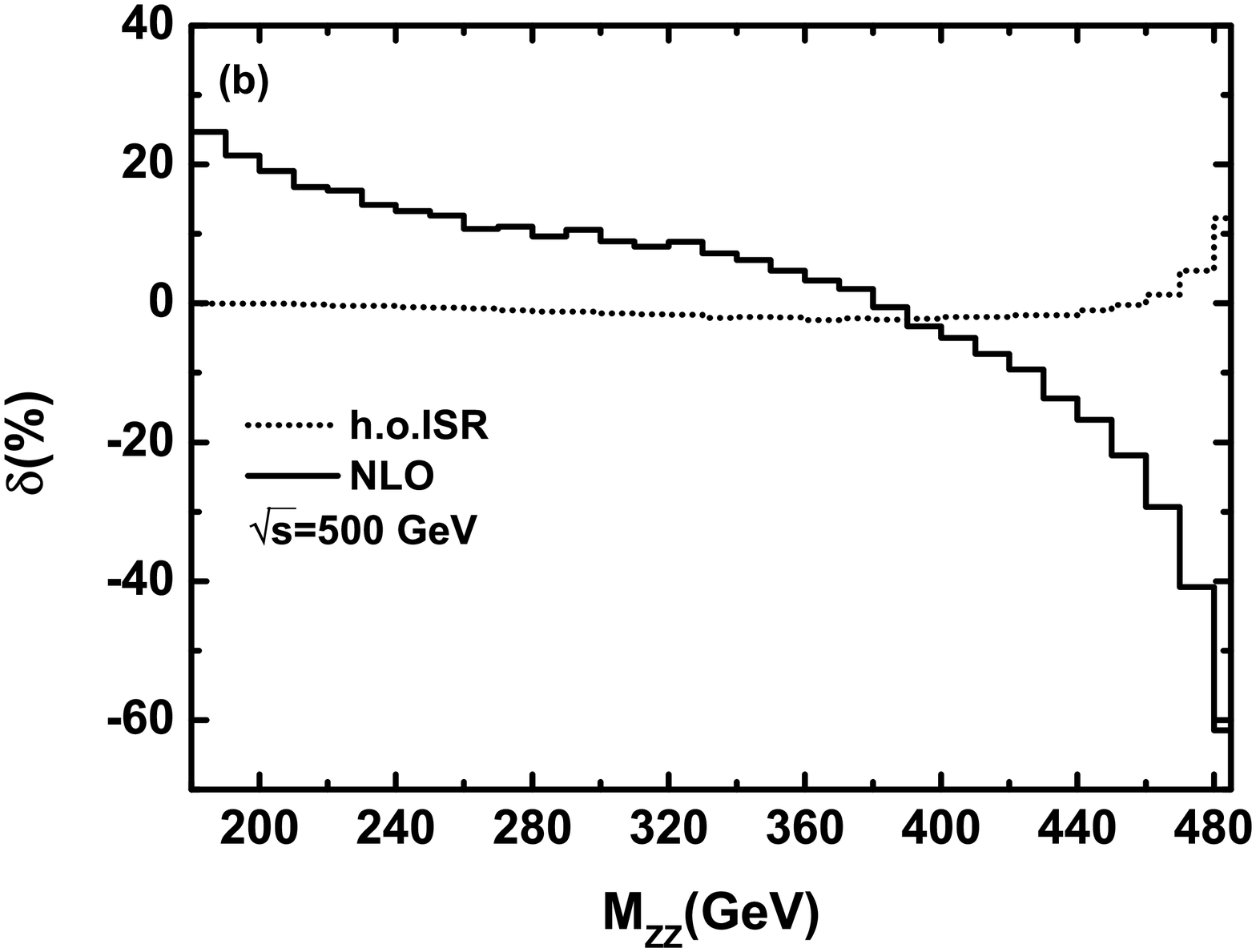}%
\hspace{0in}%
\caption{ \label{fig-mzz} (a) The LO and total EW corrected distributions of the final $Z$-pair invariant mass at the $\sqrt s=500~\GeV$ ILC. (b) The corresponding NLO EW and h.o.ISR relative corrections. }
\end{figure}

In Fig.\ref{fig-mzz}(a), we depict the LO and total EW corrected distributions of the $Z$-pair invariant mass $M_{ZZ}$. The corresponding NLO EW and h.o.ISR relative correction of $M_{ZZ}$ distribution are also presented in Fig.\ref{fig-mzz}(b). The differential cross sections of $M_{ZZ}$ are drawn in the range of $M_{ZZ}\in[2M_Z,485~\GeV]$, where the upper limit on $M_{ZZ}$ is determined by the colliding energy and the transverse momentum lower cut on the final photon. As displayed in Fig.\ref{fig-mzz}(a), the LO distribution is enhanced by the total EW correction in small $M_{ZZ}$ region, while suppressed when $M_{ZZ}>370~\GeV$. The obvious NLO EW correction in large $M_{ZZ}$ region, which can amount up to $-61.5\%$ for $M_{ZZ}\simeq 485~\GeV$, can be attributed to the Sudakov logarithms from the virtual corrections.

\par
\begin{figure}[htbp]
\includegraphics[angle=0,width=3.2in,height=2.4in]{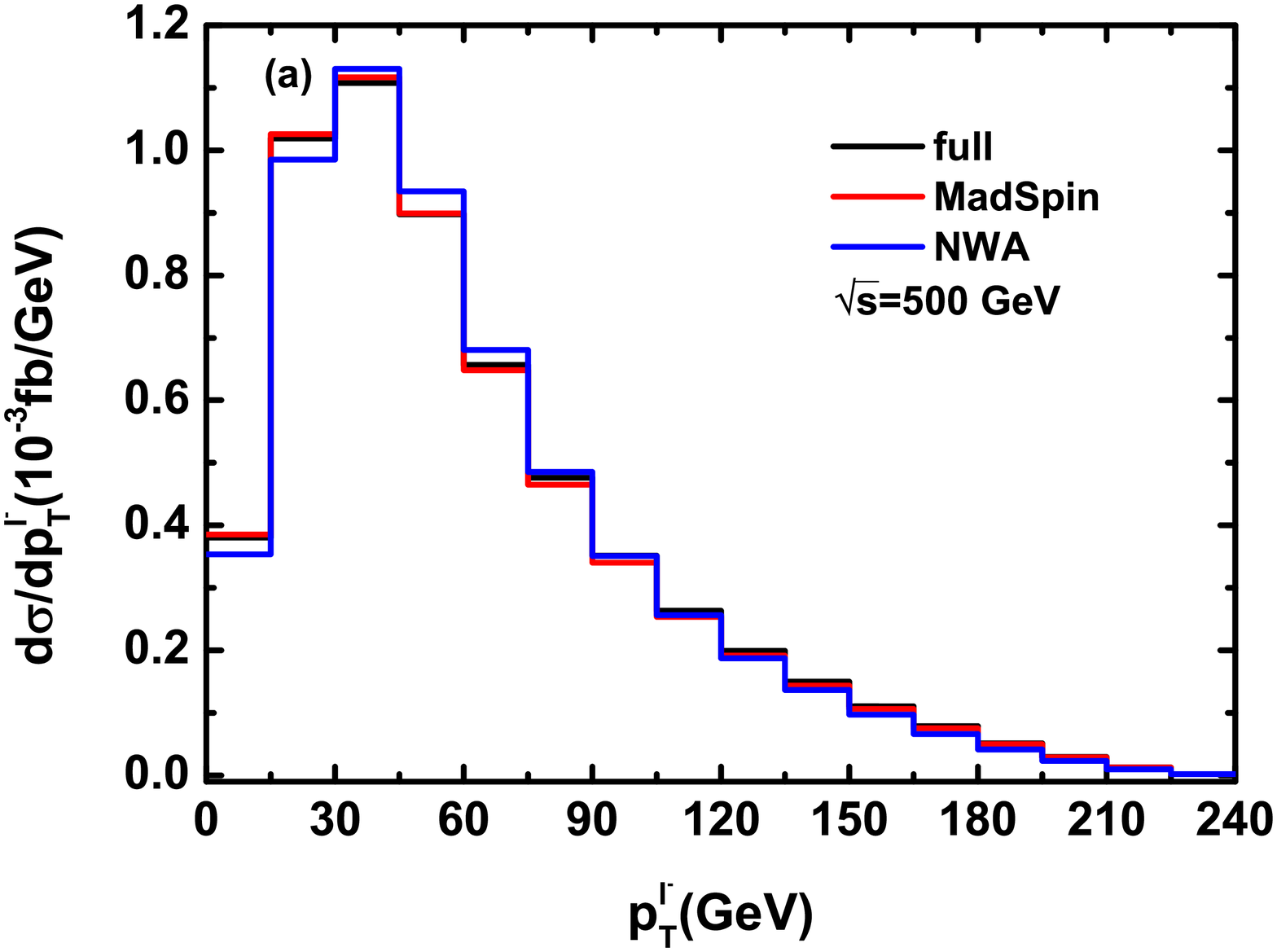}%
\hspace{0in}%
\includegraphics[angle=0,width=3.2in,height=2.4in]{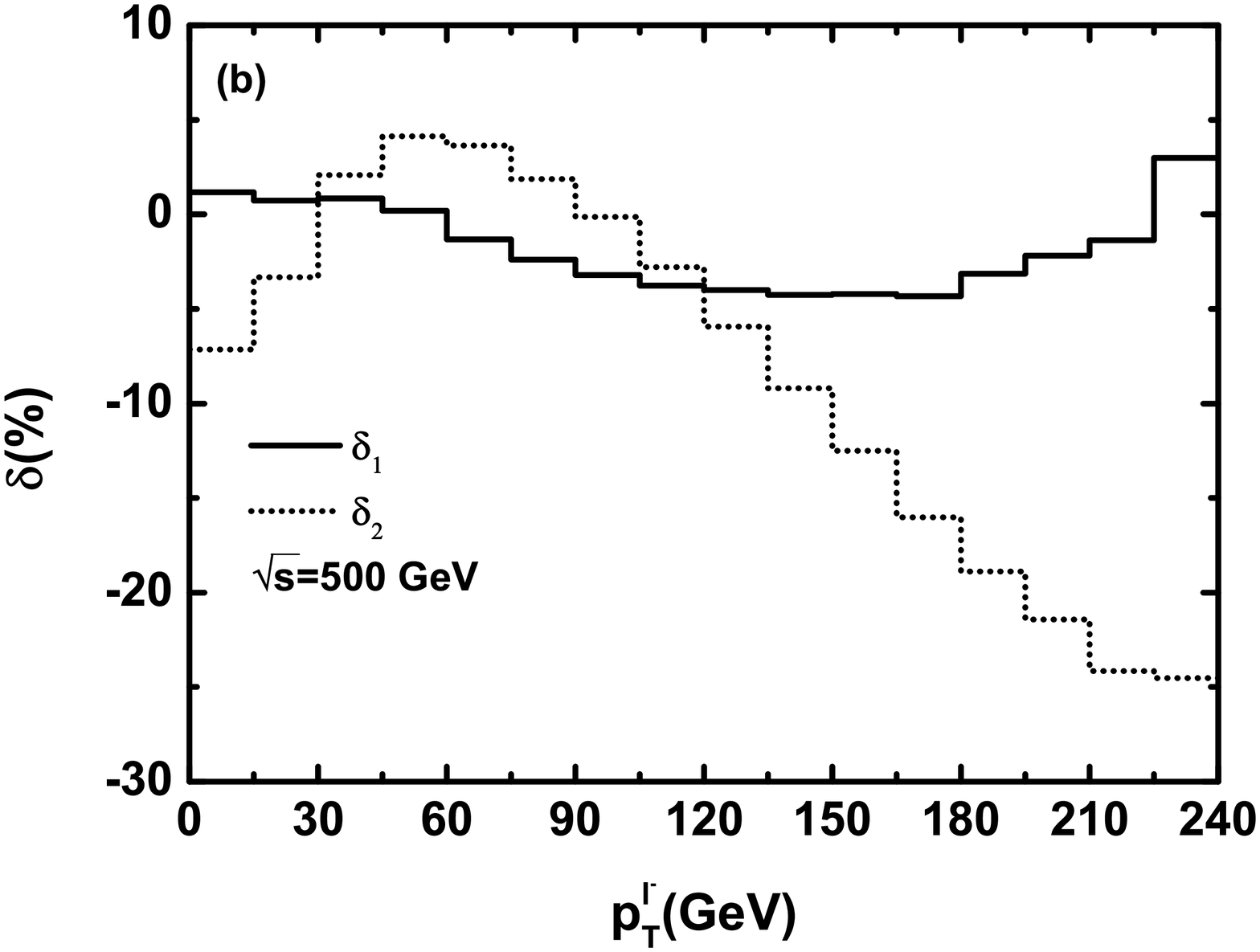}%
\hspace{0in}%
\caption{ \label{fig-com} (a) The LO transverse momentum distributions of final negative charged lepton ($l^-=e^-,\mu^-$) at the $\sqrt s=500~\GeV$ ILC by applying complete calculation, the naive NWA and {\sc MadSpin} method, i.e., $\frac{d\sigma_{{\rm full}}}{dp_T^{l^-}}$, $\frac{d\sigma_{{\rm MadSpin}}}{dp_T^{l^-}}$ and $\frac{d\sigma_{{\rm NWA}}}{dp_T^{l^-}}$. (b) The corresponding relative deviations. }
\end{figure}

Now we consider the leptonic decays of the final $Z$-boson pair and study the \eezzr$\to l_1^+l_1^-l_2^+l_2^-\gamma(+\gamma)$ ($l_1,l_2=e,\mu$) process by adopting the  following two methods within narrow width approximation (NWA) to generate the subsequent decays.
\begin{enumerate}[(1)]
\item The naive NWA method. In this method, $Z$-boson is treated as an on-shell particle and its spin information is dropped.
\item The {\sc MadSpin} method. In this method, the improved NWA is adopted and implemented in the {\sc MadSpin} package, which is part of {\sc MadGraph5}\_aMC@NLO \cite{Alwall:2014hca} and can be used to generate the heavy resonance decay taking into account the spin correlation and finite width effects to a very good accuracy, just as demonstrated in our previous papers \cite{Li:2015ura, Yong-Bai:2015xna}.
\end{enumerate}
In order to display the spin correlation effect and the accuracy of {\sc Madspin} approximation, we calculate the full amplitude for the signal process  \eezzr$\to l_1^+l_1^-l_2^+l_2^-\gamma$ at LO by taking full off-shell effects into account. In Fig.\ref{fig-com}(a), we present the LO distributions of the transverse momentum of the final negative charged lepton after $Z$-boson decays by applying both the naive NWA and {\sc MadSpin} methods separately, which are labeled as $\frac{d\sigma_{{\rm NWA}}}{dp_T^{l^-}}$ and $\frac{d\sigma_{{\rm MadSpin}}}{dp_T^{l^-}}$, respectively. The distributions are depicted by entering for each $p_T^{l^-}$ and then normalising by a factor 1/2. We also plot the same kinematic distribution for the signal process by taking full off-shell effects ($\frac{d\sigma_{{\rm full}}}{dp_T^{l^-}}$) into account in Fig.\ref{fig-com}(a). The relative deviations $\delta_1\equiv \frac{d\sigma_{{\rm MadSpin}}}{dp_T^{l^-}}/\frac{d\sigma_{{\rm full}}}{dp_T^{l^-}}-1$ and $\delta_2\equiv \frac{d\sigma_{{\rm NWA}}}{dp_T^{l^-}}/\frac{d\sigma_{{\rm full}}}{dp_T^{l^-}}-1$ are given in Fig.\ref{fig-com}(b). From these figures, we can see that the results of the naive NWA method sharply deviate from the full off-shell results (at most $25\%$), while the {\sc MadSpin} method is less deviated (less than 5\%). We see clearly that the spin correlation effect is manifested obviously in the LO $p_T^{l^-}$ distribution, and the {\sc MadSpin} program is an efficient tool in handling the spin-entangled decays of resonant $Z$-boson in an accurate way. 

\par
\begin{figure}[htbp]
\includegraphics[angle=0,width=3.2in,height=2.4in]{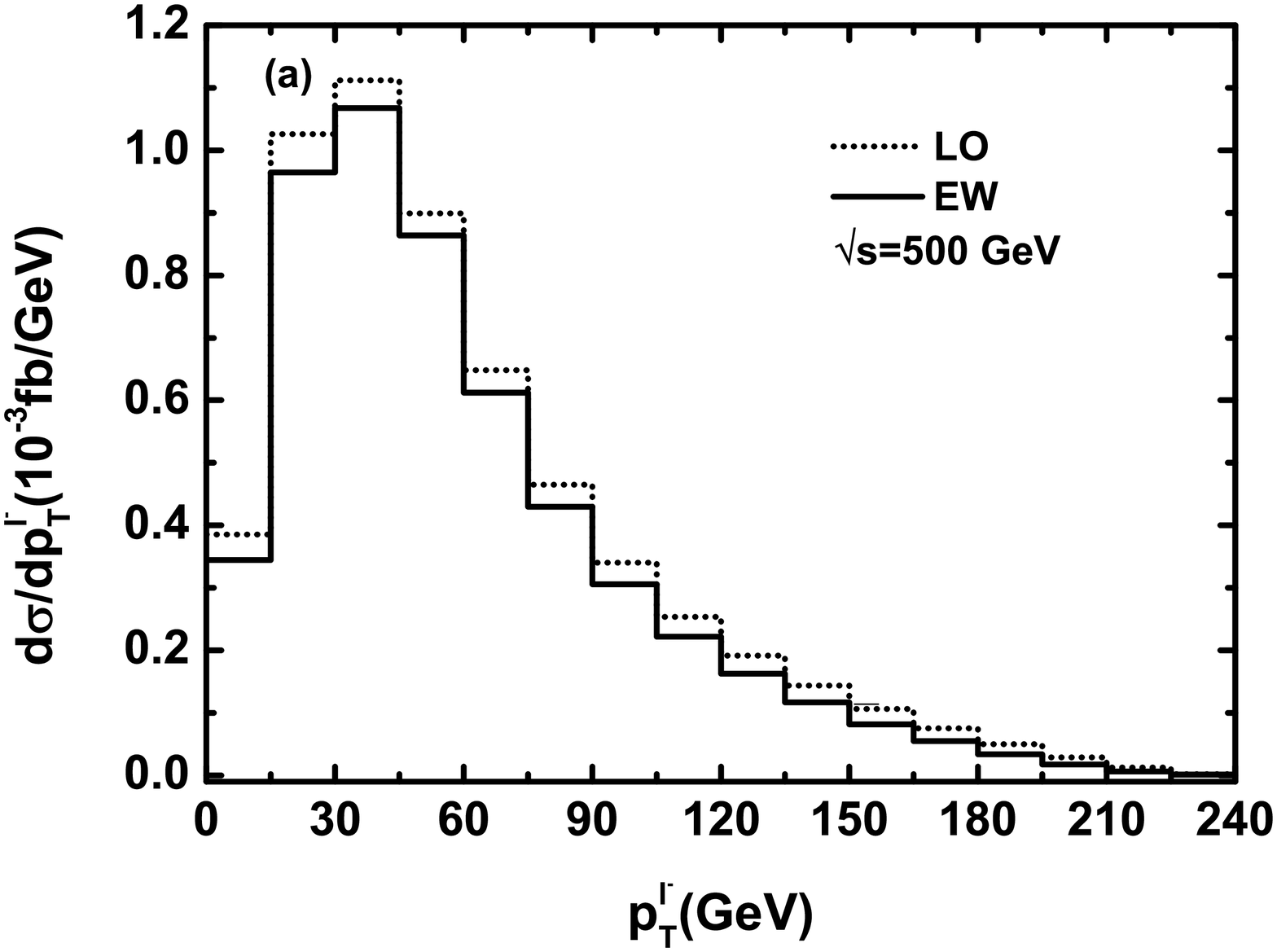}%
\hspace{0in}%
\includegraphics[angle=0,width=3.2in,height=2.4in]{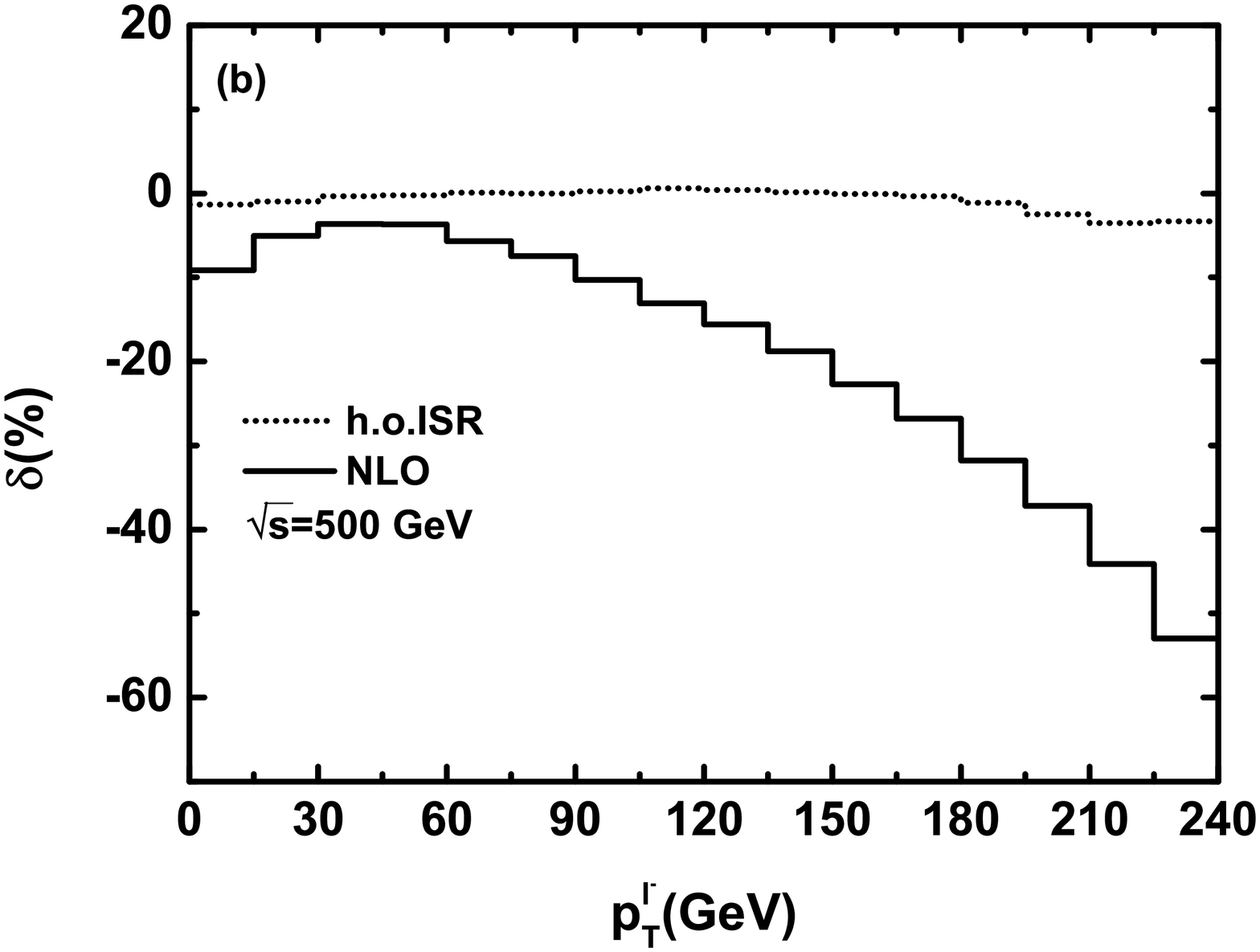}%
\hspace{0in}%
\caption{ \label{fig-pte} (a) The distributions of the LO and total EW corrected transverse momenta of final negative charged lepton ($l^-=e^-,\mu^-$) at the $\sqrt s=500~\GeV$ ILC. (b) The corresponding NLO EW and h.o.ISR relative corrections. }
\end{figure}

We present the LO and total EW corrected distributions of the final negative charged lepton transverse momentum in Fig.\ref{fig-pte}(a), and the corresponding NLO EW and h.o.ISR relative corrections in Fig.\ref{fig-pte}(b). We can see that both the LO and total EW corrected $p_T^{l^-}$ distributions reach their maxima at the position of $p_T^{l^-}\sim 30~\GeV$ and the LO distributions are suppressed significantly by EW correction in the whole plotted region.

\par
In above discussion we didn't mentioned about the QED final state radiation (FSR) from the outgoing charged leptons, which could enhanced the results in the "bare" lepton scheme in measuring final leptons due to large logarithms form $\alpha^n\ln^n(m_l^2/s)$ terms induced by the small lepton mass. But in the "dressed" lepton scheme we obtain the invariant mass $m_{ll}$ and transverse momentum $p_T^Z$ by recombining the final state leptons with radiated photons within a cone, e.g.,$\Delta R <0.1$ (called "dressed" leptons) \cite{Aad:2011gj}. Normally, final electrons are detected in calorimeters, photon recombination is automatically involved in the reconstruction from electromagnetic showers. While, muons can be observed as "bare" leptons from their tracks in the muon chambers, but in order to reduce FSR corrections we can sometimes reconstruct the observed muons as ¡±dressed¡± leptons via photon recombination  \cite{Andersen:2014efa}. In this "dressed" lepton scheme, the mass singular FSR effects vanish and the resulting cross section does not depend on the mass of charged lepton, i.e. the reconstructed lepton looks universal (at least electrons and muons). In this work we study the \eezzr$\to  l_1^+l_1^-l_2^+l_2^-\gamma(+\gamma)$ process by using the "dressed" lepton scheme.

\vskip 5mm
\section{Summary}
The \eezzr process is very important for understanding the nature of the Higgs boson and exploring the
$ZZZ\gamma$ and $ZZ\gamma\gamma$ anomalous QGCs. In this paper, we report on the full NLO EW corrections
and the h.o.ISR contributions to the $ZZ\gamma$ production at the ILC in the SM. We analyze the EW quantum effects on the total cross section and find that both the NLO EW and total EW corrections suppress the LO cross section when $\sqrt s$ goes up from $200~\GeV$ to $1~\TeV$. Due to the Coulomb singularity effect, the NLO EW relative correction is very large near the threshold (e.g., $\delta_{NLO}=42.72\%$ with $\sqrt s =200~\GeV$). The h.o.ISR effect beyond $\cO(\alpha)$ is also distinct near the threshold (e.g., the relative correction is $11.81\%$ when $\sqrt s =200~\GeV$), while becomes small at the high colliding energy region. We provide the LO and total EW corrected $p_T^Z$, $p_T^\gamma$ and $M_{ZZ}$ distributions and investigate the corresponding NLO EW relative correction and h.o.ISR relative correction. We find that the h.o.ISR effect becomes notable near the maximum value of $p_T^Z$, $p_T^\gamma$ and $M_{ZZ}$. We also investigate the leptonic decays of the final $Z$-boson pair by adopting the {\sc MadSpin} program to include the spin correlation effect and find that the LO $p_T^{l^-}$ distributions are suppressed in the whole plotted region. Ascribed to the Sudakov effect, the NLO EW correction becomes very large with the increment of these kinematic variables. We conclude that both the h.o.ISR and the NLO EW corrections are worth being taken into account in precision measurement of the $ZZ\gamma$ production at the ILC.

\vskip 5mm
\section{Acknowledgments}
We gratefully thanks to Li Wei-Hua for many helpful discussions. This work was supported in part by the National Natural Science Foundation of China (Grant No.11405076, No.11347101, No.11275190, No.11375171 and No.11535002), the Startup Foundation for Doctors of Kunming University of Science and Technology (Grant No.KKSY201556046, No.KKSY201356060), and the Science Foundation of Yunnan Provincial Education Department (Grant No.2014Y066).

\end{document}